\documentclass[aps,pr,twocolumn]{revtex4}  
\usepackage{epsfig}
  
\begin{document}  
  
\begin{flushleft}  
LYCEN 2002-38  
\end{flushleft}  
  
\title{Hydrodynamics with spontaneous symmetry breaking:
application to relativistic heavy ion collisions}  
\author{Y. Lallouet, D. Davesne and C. Pujol}  
\address{IPN Lyon, 43 Bd du 11 Novembre 1918, F-69622 Villeurbanne Cedex}   
\date{\today}
\begin{abstract}  
In this paper we apply hydrodynamics for systems with continuous broken
symmetries to heavy ion collisions in the framework of (1+1) dimensional Bjorken
model. The temperature profile with respect to proper time determined in that
context exhibits no differences with the ideal fluid. On the contrary, it is
shown that the profile obtained when M\"{u}ller-Israel-Stewart second
order theory of dissipation is included on top of standard hydrodynamics
indicates a slower cooling of the system.
\end{abstract}  
\maketitle  
  
\section{Introduction}  
Relativistic hydrodynamics is a very useful tool in the field of heavy ion
collisions when one wants to reproduce the space-time evolution of the hot and
dense matter created in such events. Upon the hypothesis of local equilibrium,
which can be questionable for such small and not long-lived systems, this
approach directly gives in terms of usual quantities such as temperature,
pressure..., an intuitive description of what happens during the whole evolution
of the system until the freeze-out takes place. Moreover, it is usually assumed
that the matter under consideration is almost perfect, so that standard
hydrodynamics can be applied directly. However, the hot and dense matter
produced is mainly constituted of pions and, with the exception of \cite{son},
no attention has been drawn to the fact that the breaking of chiral symmetry can
affect the theory of hydrodynamics itself. One of the aims of this paper is thus
to give in a simple approach a quantitative estimate of this modification. More
precisely, we will determine the temperature profile with respect to time in the
framework of the (1+1) dimensional boost-invariant Bjorken model with and
without the effect of the spontaneous breaking of the symmetry. In a second step
we also introduce the dissipation process and compare the previous results with
those obtained when M\"{u}ller-Israel-Stewart second order theory of dissipation
is included on top of standard hydrodynamics.

\section{Hydrodynamics equations for chiral fluid}   

Following \cite{son}, the hydrodynamics degrees of freedom in a chiral fluid
with chiral $SU(2)$ symmetry spontaneously broken are the densities associated
to conserved quantities, namely entropy density $s$, momentum density $T^{0i}$,
baryonic number density $n$, left and right-handed charge densities which can be
written as $SU(2)$-matrices $\rho_L\equiv\rho_L^i\tau_i/2$ and
$\rho_R\equiv\rho_R^i \tau_i/2$ and finally the variables associated to the
Goldstone modes introduced in the theory {\it via} a $SU(2)$-matrix $\Sigma
\equiv e^{i \vec \tau .\vec \pi / f_{\pi}}$. If we focus on the central rapidity
region of a heavy ion collision, we can neglect in a first step the baryonic
density. Thus, the hydrodynamic equations taken from \cite{son} become:

\begin{equation}  
\partial_\mu T^{\mu\nu}=0  
\label{tmunu}
\end{equation}  
\begin{equation}  
i\partial_\mu((f_t^2-f_s^2)u^\mu u^\nu\Sigma\partial_\nu\Sigma  
^\dagger+f_s^2\Sigma\partial^\mu\Sigma^\dagger) = 0  
\label{sigma}
\end{equation}  
where the energy-momentum tensor $T^{\mu\nu}$ is given by $T^{\mu\nu}=(\epsilon
+ p)u^\mu u^\nu - p g^{\mu\nu} + f_s^2/4\hbox{tr}(\partial^\mu\Sigma\partial^\nu
\Sigma^\dagger + \partial^\nu\Sigma\partial^\mu\Sigma^\dagger)$.

In the above equations, $f_t$ ($f_s$), which can be interpreted as the
temporal (spatial) pion decay constant, are actually not constants but have a
non-trivial temperature dependence. Nevertheless, in order to have an estimation
of the effects induced by the presence of the spontaneous breaking of the chiral
symmetry, we will adopt two simple approaches. The first one, for the sake of
simplicity, is to consider that $f_t = f_s = f_{\pi} = 93$ MeV. Of course, at
finite temperature, this is not true and, for instance, for small temperatures,
one knows that $f_t^2$ and $f_s^2$ differ from $f_{\pi}^2$ by a factor
proportional to $T^2$ while the correction for $f_t^2 - f_s^2$ is of order $T^4$
so that, to first order, $f_t^2 = f_s^2 = f_{\pi}^2 (1-T^2/(6f_{\pi}^2))$. This
second approach, more realistic, is based on this formula, extended to larger
temperatures.

\section{The Bjorken picture}   

Even if several important improvements concerning the geometric description of a
heavy ion collision have been done since the pioneering paper of Bjorken
\cite{Bjorken}, we will use here this simplified version. The Bjorken picture of
a collision is based on the hypothesis that initial conditions are invariant
under Lorentz transformations throughout the central rapidity plateau. This
implies that, for a one-dimensional flow in the $z$-direction, all the
quantities depend actually not on the time $t$ and the position $z$ but only on
the proper time $\tau = \sqrt{t^2-z^2}$. The space-time evolution of the system
can be then represented by hyperbola $\tau = cst$. If the hydrodynamic
velocity $u^{\mu}$ is taken as $u^{\mu} = x^{\mu}/\tau = (t/\tau,0,0,z/\tau)$,
the hydrodynamic equations
\begin{equation}  
\partial_\mu T^{\mu\nu}=0  
\end{equation}  
with $T^{\mu\nu}=(\epsilon + p)u^\mu u^\nu - p g^{\mu\nu}$ lead for an
ultra relativistic gas to the following differential equations
\cite{csernai,CYW}:
\begin{equation}
\frac{\partial s}{\partial \tau}+\frac{s}{\tau}=0
\end{equation}
\begin{equation}
\frac{\partial T}{\partial \tau}+\frac{1}{3}\frac{T}{\tau}=0
\end{equation}
\begin{equation}
\frac{\partial \epsilon}{\partial \tau}+\frac{4}{3}\frac{\epsilon}{\tau}=0
\end{equation}
whose solutions give the following scaling behavior:
\begin{equation}
s(\tau)=s(\tau_{0})
\frac{\tau_{0}}{\tau}
\end{equation}
\begin{equation}
T(\tau)=T(\tau_{0})
\left (
\frac{\tau_{0}}{\tau}
\right )^{{1}/{3}}
\end{equation}
\begin{equation}
\epsilon(\tau)=\epsilon(\tau_{0})
\left (
\frac{\tau_{0}}{\tau}
\right )^{{4}/{3}}.
\end{equation}
And, since $p=\epsilon /3$:
\begin{equation}
p(\tau)=p(\tau_{0})
\left (
\frac{\tau_{0}}{\tau}
\right )^{{4}/{3}}.
\end{equation}

\section{Results}   

We first take into account the effect of the spontaneous breaking of chiral
symmetry where no renormalization of the pion decay constant occurs, i.e.
$f_t = f_s = f_{\pi} = 93$ MeV. We thus obtain from the hydrodynamic equations
(\ref{tmunu}),(\ref{sigma}) in the first non-vanishing order of the pion field~: 
\begin{equation}
\frac{\partial s}{\partial \tau}+\frac{s}{\tau}=0
\end{equation}
\begin{equation}
\frac{\partial T}{\partial \tau}+\frac{1}{3}\frac{T}{\tau}=0
\end{equation}
\begin{equation}
\frac{\partial \epsilon}{\partial \tau}+\frac{Ts(\tau_{0})\tau_{0}}{\tau^{2}} -
\frac{1}{\tau}\sum_{i=1}^3 
\left(
\frac{\partial
{\pi^i} }
{\partial\tau}
\right )^{2}
=0 
\end{equation}
\begin{equation}
\left(
\frac{\partial}
{\partial\tau}+\frac{1}{\tau}
\right)
\frac{\partial{\pi^i}}{\partial \tau}=0. 
\label{pifts}
\end{equation} 
and from the explicit expression of the pressure $p$ given in \cite{son}:
\begin{equation}
p=aT^4+\frac{1}{2}
\sum_{i=1}^3
\left (
\frac{\partial
{\pi^i} }
{\partial\tau}
\right )^{2}
\label{pressfts} 
\end{equation}
We see that there is no modification of the equation for the entropy density and
for the temperature, that is the scaling behavior is exactly the same
$s(\tau)=s(\tau_{0})\tau_{0}/\tau$ and $T(\tau)=T(\tau_{0}) \left( \tau_{0}/
\tau \right)^{{1}/{3}}$. On the contrary, for the energy density and for the
pressure, one can observe a modification due to a term proportional to $\partial
{\pi^i}/\partial \tau$ which can be determined from the last equation
(\ref{pifts}). The solution for the time derivative is $\partial{\pi^i}/\partial
\tau = A/\tau$ where $A$ is a constant whose value is fixed by initial condition
for $\tau = \tau_0$. If we now report in the above equations, we can solve
analytically the resulting system to obtain:
\begin{equation}
\epsilon(\tau)=\epsilon(\tau_{0})
\left (
\frac{\tau_{0}}{\tau}
\right )^{{4}/{3}}
- \frac{3}{2}\frac{A^2}{\tau^2_0}
\left(\frac{\tau_0}{\tau}\right)^2
\label{enerftsa}
\end{equation}
\begin{equation}
p=p(\tau_{0})
\left (
\frac{\tau_{0}}{\tau}
\right )^{{4}/{3}}+\frac{3}{2}\frac{A^2}{\tau^2_0}\left(\frac{\tau_0}{\tau}
\right)^2.
\label{pressftsa} 
\end{equation}

In order to fix the initial condition related to the value of $A$, let us now
consider the density matrix $\rho = \exp ^{-\beta H}$. The part of the
Hamiltonian relevant for chiral hydrodynamics degrees of freedom can be taken
from \cite{son}. An explicit calculation with Bjorken modelisation of the
collision gives:
\begin{equation}
\rho=\exp\left( -\beta\int\hbox{d}^3xT^{00}\right)
\propto\exp\left(-\frac{\beta}{2}\int\hbox{d}^3x
\sum_{i=1}^3
\left (
\frac{\partial
{\pi^i} }
{\partial\tau}
\right )^{2}\right).
\label{densi} 
\end{equation}
Assuming isospin symmetry, we then have: $\rho\propto~\exp\left(-\frac{3\beta}
{2}\int\hbox{d}^3x \left( \frac{A}{\tau} \right) ^{2}\right)\label{densi}$. For
$\tau = \tau_0$, we thus immediately see that $\rho$ is Gauss distributed.
Actually, if we call $V_0$ the initial volume in the rest-frame of the two
colliding nuclei and $\gamma _{cm}$ the Lorentz factor needed to go from that
frame to the laboratory frame, we have:
\begin{equation}
\rho(\tau_0)
\propto\exp\left(-\frac{3\beta}{2}\frac{V_0}{\gamma_{cm}}
\left (
\frac{A}{\tau_0}
\right )^{2}\right).
\label{densio} 
\end{equation}
We can easily estimate $V_0$ by geometric arguments: $V_0=\pi R^2 2\tau_0$ where
$R$ is the radius of the colliding nuclei. If we now put typical numbers on all
these constants: $R\simeq 5$ fm, $\tau_0=1$ fm, $\gamma_{cm}\simeq 10$ and
$T(\tau_0) = 200$ MeV, we can directly estimate the effect of the terms
associated to the spontaneous symmetry breaking since $<A^2>$ is simply related
to the Gaussian distribution: $<A^2> =T(\tau_0)\tau_0\gamma_{cm}/
(6\pi R^2)$. Numerically, using these inputs, we obtain a correction of order
$\simeq $1\% with respect to the perfect case.

\begin{figure}[H]
  \begin{center}
    \mbox{
          \epsfysize=7.4cm
          \epsfbox{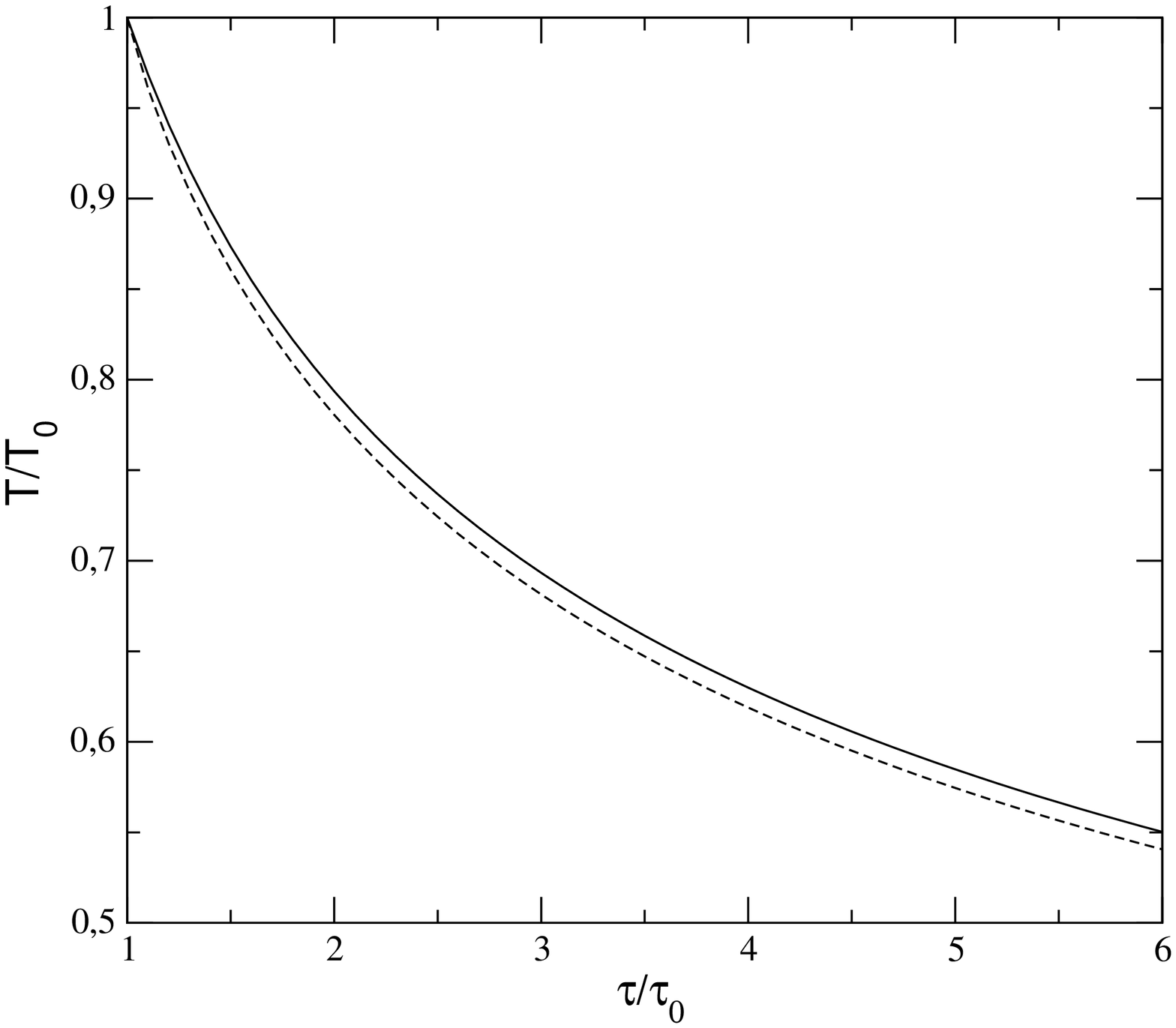}}
  \end{center}
  \caption{Temperature profile with respect to proper time for two cases:
  without (solid line) and with (dashed line) effect of symmetry breaking.}
  \label{graphbreak}
\end{figure}

We now introduce the explicit chiral perturbation $T$-dependence for $f_t$ and
$f_s$, i.e. $f_t^2 = f_s^2 = f_{\pi}^2 (1-T^2/(6f_{\pi}^2))$. The evolution of
the temperature is then governed by the following equation~:
\begin{equation}
4a\tau T^3
-\frac{<A'^2>}{\tau\left(1-T^2/(6f_{\pi}^2)\right)^2}\frac{T}{6f_{\pi}^2} =
s(\tau_0)\tau_0 
\end{equation}
where now $<A'^2> = <A^2>(1-T_0^2/(6f_{\pi}^2))$. The numerical results are
presented on figure \ref{graphbreak}. We can observe that the correction, which
induce a faster cooling of the system is of order 5\%.

Even if the approach can be considered too simplistic, it is not reasonably
expected that an improved treatment can modify noticeably the order of
magnitude. One can thus conclude that the effect due the spontaneous breaking of
chiral symmetry is quite negligible and without any experimental consequences. 

\section{Dissipative effects}

In the previous part we have seen that it is possible to neglect the effect of
the breaking of the symmetry on the hydrodynamics theory itself, that is we can
use "normal" hydrodynamics. Nevertheless an important question which has
already been addressed in \cite{murongaWinter} is the effect of dissipation on
top of this model of perfect fluid.

There are two non equivalent possibilities to take into account dissipation.
First, one can use the usual formalism and just say that when an inhomogeneity
appears in a system, the latter reacts and a flow proportional to the gradient
develops itself in order to restore equilibrium. This is first order theory.
However it is well-known that this approach presents instabilities and that it
is necessary to consider a relaxation time for the flow. This goal is achieved
through M\"{u}ller-Israel-Stewart second order theory. In the following we will
explore both methods.

In \cite{murongaWinter}, the author showed that only the dissipative effects
related to the shear viscosity $\eta$ play a role for the temperature profile in
the (1+1) dimensional Bjorken model. The main physical reason is that bulk
viscosity vanishes for an ultra relativistic gas. He took the shear viscosity
from \cite{Welke} where the temperature dependence for $\eta$ is $\eta=\pi
f^4_{\pi}/(8T)$. Then, the differential equations to be solved for the energy
density $\epsilon = 3aT^4$ are: 
\begin{equation}
\frac{\partial\epsilon}{\partial\tau}=-\frac{4}{3}\frac{\epsilon}{\tau}+
\frac{\Phi}{\tau}
\label{evisc}
\end{equation}
where $\Phi$ is related to the stress tensor (see \cite{murongaWinter} for
further details) and $\Phi = 0$ for the ideal (perfect) fluid, $\Phi=\frac{4}{3}
\frac{\eta}{\tau}$ for the first order and $\tau_{\pi}\frac{\partial\Phi}
{\partial\tau} = -\Phi+\frac{4}{3}\frac{\eta}{\tau}$ for the second order theory
discussed before. In the last equation $\tau_{\pi}$ is the relaxation time
associated to $\Phi$.

\begin{figure}[H]
  \begin{center}
    \mbox{
          \epsfysize=7.4cm
          \epsfbox{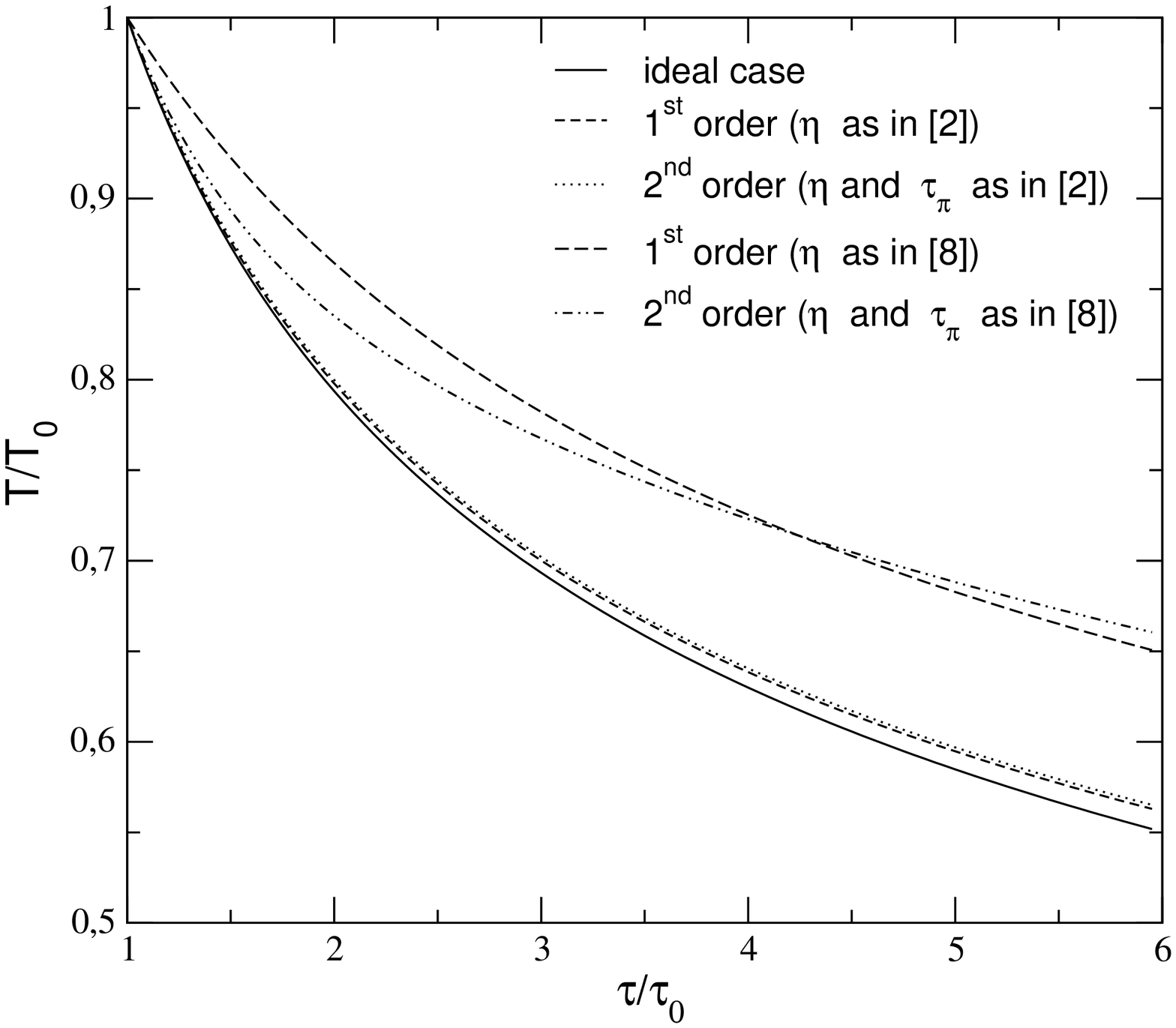}}
  \end{center}
  \caption{Temperature profile with respect to proper time for the perfect
  fluid (solid line), first order and second order theories.}
  \label{graphvisc}
\end{figure}

We now use (\ref{evisc}), using the same inputs as \cite{murongaWinter}, namely
massless pions and current algebra cross section. We can see on figure
\ref{graphvisc} that the correction due to viscosity compared to the ideal case
is totally negligible not only for the (non-physical but usual) first order but
also for the second order. Our results are different from those presented in the
original paper \cite{murongaWinter}. The difference lies in fact in a small
computational mistake in \cite{murongaWinter}. The main point is that, contrary
to what is claimed in \cite{murongaWinter}, there are, in this simple approach,
no appreciable effects. If we would stop here, we could say, as in the first
part dedicated to the breaking of chiral symmetry that a realistic description
of a pion gas can be achieved with the perfect fluid model.

However, we stress that this conclusion only holds because of the hypothesis
used in \cite{murongaWinter}. The temperature dependences of the viscosity
$\eta$ and of the relaxation time $\tau_{\pi}$ are only valid for massless pions
and when the cross-section for $\pi-\pi$ interaction is given by current
algebra. If we take instead the realistic case of massive pions and the
experimental cross-section, these dependences are completely different. For
instance, $\eta$ increases with temperature \cite{Welke,moi} instead of
decreasing. With this realistic $T$-dependence for the viscosity and the
relaxation time for the pions taken from \cite{moi}, we obtain the curves
plotted on figure \ref{graphvisc}. We can now observe that the first and the
second order are still close but they are now around 20\% above the ideal
fluid case. This means that dissipative effects imply a slower cooling of the
system and should be incorporated in the simulations of heavy ion collisions.

\section{Conclusion}  

In the first part of this paper we have examined in a simple model an
application of hydrodynamic equations with spontaneous breaking symmetry. In
order to get an estimate of the magnitude of the effects, we have used the one
dimensional Bjorken description of a heavy ion collision. We have shown that the
model of the perfect fluid is almost correct~: the effect of the breaking of the
symmetry is to give a correction of about 1\% when no $T$-dependence of the
pion decay constant is taken into account and a few percents when we consider a
crude $T$-dependence. In a second part we have considered an other deviation
from the perfect fluid model, namely the dissipation. The dissipative effects,
when taken into account in a simple description \cite{murongaWinter} are not so
important as claimed in this reference but represent actually a few percents. On
the contrary, we have obtained sizable effects with a correct $T$-dependence for
the shear viscosity and the shear flow relaxation time. For example, for an
initial temperature of 200 MeV, the temperature goes from about 110 MeV in the
ideal case to 140 MeV with our dissipative effects for $\tau/\tau_0 \simeq 6$.
The whole conclusion of this paper is that the temperature profile generated
from the perfect fluid approach and used in many simulations is almost correct
if one takes into account only the effects of symmetry breaking. On the
contrary, if we consider dissipative effects the profile can be substantially
modified, that is we have a slower cooling of the pion gas.
 
\vglue 1 true cm

Acknowledgments~: We thank M. Ericson, Emeritus Professor at University Claude
Bernard and G. Chanfray for constant help and enlightening discussions. We also
thank J. Serreau for many discussions.

\end{document}